# An Experimental Evaluation of Computational Techniques for Planning and Assessment of International Interventions

*Alexander Kott, Army Research Laboratory, Adelphi MD*

*Jeff Hansberger, Army Research Laboratory, Adelphi MD*

*Edward Waltz, BAE Systems, Arlington VA*

*Peter Corpac, Science and Technology Associates, Arlington VA*

## Abstract:

We describe the experimental methodology developed and employed in a series of experiments within the Defense Advanced Research Projects Agency (DARPA) Conflict Modeling, Planning, and Outcomes Exploration (COMPOEX) Program. The primary purpose of the effort was development of tools and methods for analysis, planning and predictive assessment of plans for complex operations where integrated political-military-economic-social-infrastructure and information (PMESII) considerations play decisive roles. As part of the program, our team executed several broad-based experiments, involving dozens of experts from several agencies simultaneously. The methodology evolved from one experiment to another because of the lessons learned. The paper presents the motivation, objectives, and structure of this interagency experiment series; the methods we explored in the experiments; and the results, lessons learned and recommendations for future efforts of such nature.

## Complex Interagency Operations

Interagency decision-making is particularly significant in complex international operations, such as stability and peace operations. In order to synchronize effectively appropriate elements of national capabilities, interagency teams need to consider how to coordinate actions of multiple agencies to achieve a coherent set of desired effects. Complex interagency operations are characterized by:

- Situations that involve highly interconnected dynamic and adaptive political, social, economic, infrastructure and information systems, as well as the formal militaries and unstructured forces (insurgencies, criminal entities, etc.) operating within that environment. Such systems of systems are often characterized by uncertainty and instability – and are inherently unpredictable.

- Necessity to plan, adapt and orchestrate all elements of national power to effectively perform shaping, deterrence, containment, defeat or restoration; this requires the coordination of interagency contributors, and an integrated plan that represents the whole of government.

Iraq, Afghanistan, Somalia, Bosnia, and Kosovo are all examples of complex international operations. Such operations present the necessity to plan, adapt, and orchestrate the appropriate elements of national power to effectively perform shaping, deterrence, containment, defeat, or restoration. To accomplish this, interagency teams require demanding capabilities:

- *A means to represent rapidly changing situations* – Interagency teams require a systems understanding of an evolving situation to provide insight into structural characteristics and behavioral dynamics. Systems considerations allow leaders and their staffs to consider a broader set of options to create desired effects while avoiding undesired effects (TRADOC 2008). This systems view also provides a shared understanding across the team.

- *Coordination of interagency contributors* – Teams also require a means to coordinate their shared vision of alternatives; component plans from Department of Defense (DoD), Department of State, United States Agency for International Development (USAID), and other agencies must be developed, integrated into a whole and evaluated for their combined effects.

- *Dynamic analysis of the potential effects of plans* – Behavioral analysis, performed by games, exercises or simulations to predict potential effects (consequences), stimulates in-depth thought about the operation, causing the planning staff to consider the underlying dynamics of target systems - gaining insights that otherwise might not have occurred.

- *Production of an integrated plan* – Teams require a means to develop and represent coordinated plans that are integrated, yet present information in the perspective and language of each agency. DoD, for example, focuses on the time-sequencing of intense activities (synchronization matrix perspective), while Department of State and USAID organizations focus on allocations to standard aid project categories (a budget planning perspective).

These situations present incredibly complex and difficult problems to be solved. This paper uses the DARPA COnflict Modeling, Planning and Outcomes Experimentation (COMPOEX) program (Waltz 2008; Kott and Corpac 2007) to present an experimentation methodology for whole of government planning and wargaming of complex international operations. It provides an understanding of the experimental methods, tools, results and lessons learned.

## COMPOEX Program and Approach

The COMPOEX program (originally known as Integrated Battle Command (IBC)) developed decision support tools to aid decision-makers in planning, visualizing and executing whole-of-government major operations. It begun in 2004 as a collaborative effort between DARPA and Joint Forces Command (JFCOM) to develop technologies that could enhance the capability of leaders and staffs to plan and execute major operations in a complex environment (Fig. 1).

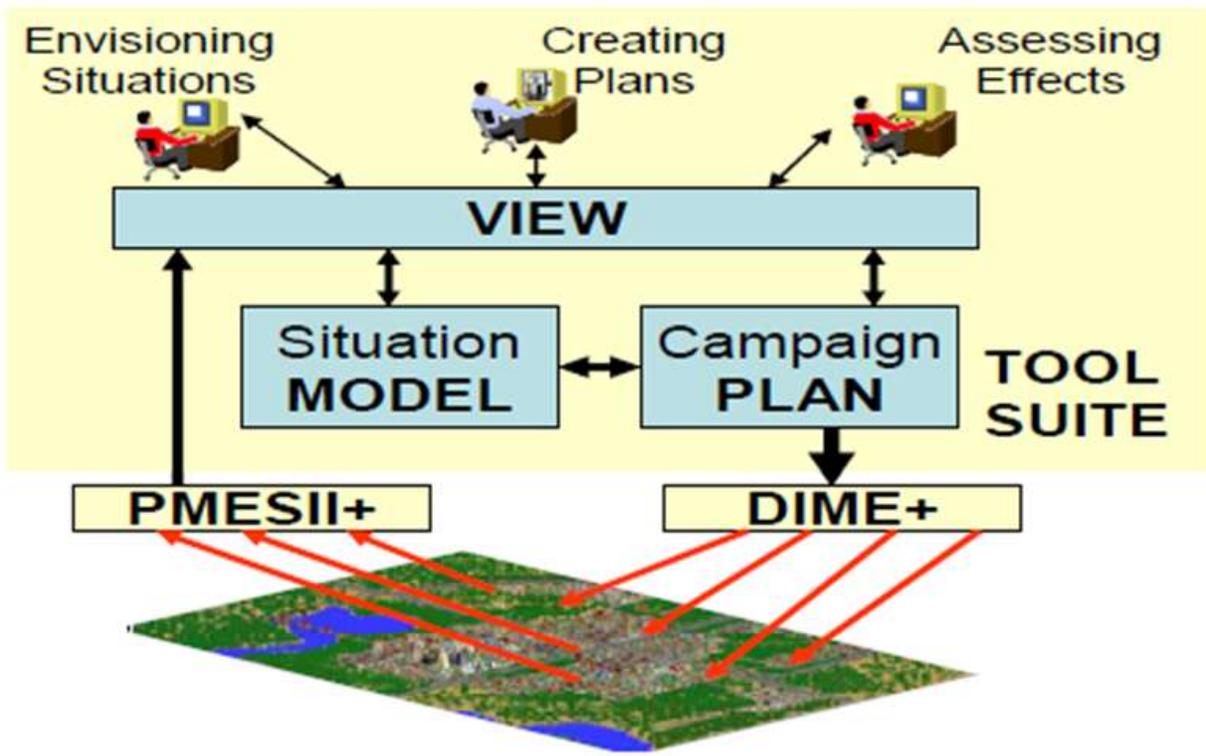

*Fig. 1. Experiments focused on tools and techniques that help leaders and staffs to plan and execute major operations in a complex environment.*

In the vision of the program, military and civilian leaders must jointly visualize, understand and effectively operate in the complex Political, Military, Economic, Social, Information, Infrastructure (PMESII) environments and employ a whole-of-government approach to planning and executing major operations.  It was important to explore possible actions to determine the range of plausible effects, and to plan long-range operations, encompassing various lines of effort to achieve national objectives (Honey et al. 2003). Interagency experimentation was critical in both evaluating the new planning tools as well as developing new methods to best utilize the emerging technology in whole-of-government planning and execution.

The COMPOEX program provided a variety of tools—and associated methods--to assist in planning and executing whole-of-government operation plans.  This diversity of tools, methods and perspectives contrasts strongly with approaches where planning and assessment are focused on one aspect of an operation, such as a strictly military engagement of enemy forces (Kott et al. 2002). The tools included a family of interconnected complex multi-resolution models (Waltz 2008) to represent the operational environment. We introduce these tools later in this paper.

## The Series of Experiments

The COMPOEX program proceeded through a series of extensive experiments. Each experiment explored COMPOEX tool capabilities and the corresponding methods, and evaluated their impact on interagency staffs and whole-of-government planning processes. The experiments built on previous ones to expand proven tools and methods, and to evaluate new capabilities.

The first in the series of experiments, called Effects Identification, asked an interagency team to evaluate the effectiveness of two different sets of tools. This experiment took place in May 2006, 12 months after the beginning of the COMPOEX program, took two full weeks to complete, and involved four senior military leaders, five senior government agency leaders, and thirty staff and support personnel. Several teams, led by experienced senior interagency and military leaders, planned international operations, such as prevention of civil war in a conflict-torn country, elimination of militia threats, or a post-war reconstruction. A typical team included a former senior State Department, USAID or National Security Council leader, a senior military leader and three staff members with military, state or justice department experience working on an interagency staff planning current operations. Two teams that used two different sets of COMPOEX tools and a control group that used conventional tools independently planned three different scenarios. The teams rotated tool sets at the end of each of the three scenarios.

The teams explored the range of available options: actions against different nodes, such as key individuals or organizations; modification of the strength of the actions, such as funding multiple projects supporting a faction or increasing military operations; changes to the timing and sequence of action(s), and synchronizing multiple actions. An example would include a security action against a disruptive militia while using diplomatic pressure to isolate them from other factions and organizations, while also funding reconstruction projects in friendly areas and an extensive information campaign to explain the actions to affected populations and others. Each team presented an outbrief that included their recommended courses of action and expected impacts on the situation and alternative actions considered and discarded, with supporting reasons.

Evaluation focused not on the quality of the resulting solutions, but rather on interagency team methods and interactions with the tool suites. The experiment evaluation showed that one set of tools was able to explore more actions and identify more significant outcomes than the other. Only that set of tools was used for further development and experimentation in the COMPOEX program.

Experiment 2, called Domain Visualization, used two teams, one consisting of members with predominantly military backgrounds, and the other predominantly civilian but also including a few military members. This experiment took place in January 2007, over a week, supported by five interagency and five military staff, with ten support personnel. The team members had military, State Department, USAID, National Security Council, Department of Justice and rule of law backgrounds. The objective of the experiment was to determine the effectiveness of the PMESII data visualization in helping teams understand complex domain information.

An added benefit of the experiment was to highlight differences between military and civilian planning teams' understanding of the problem, methodology for developing solutions and approaches to meeting experiment objectives. We found that the military-dominated team was task-oriented and focused on providing the required reports on schedule. The civilian interagency team spent the majority of their time looking at the problem from a broad variety of perspectives, developing a method for solving it, but was unable to produce the full experiment deliverables within the allotted time. Military members of the civilian-dominated team were uncomfortable with the perceived lack of task-orientation.

Experiment 3, called Operation Planning, focused rather narrowly on effectiveness and usability of COMPOEX tools and methods in developing an interagency operation plan. This experiment took place in March 2007, took three days and involved five interagency and five military retired leaders with military, State Department, USAID and National Security Council experience. This and subsequent experiments also identified significant differences: the military had more extensive planning experience and experimentation experience.

Experiment 4, called Parallel Planning, produced a hypothetical plan for an actual ongoing operation. The efforts proceeded in parallel with a similar planning effort by an actual planning staff that supported the actual operation. The mission was to formulate a range of diplomatic, information, military, and economic actions for obtaining ten specifically named effects.

This experiment started with a three-day workshop in April of 2007 that brought together notable subject matter experts from the US and two foreign militaries, State Department, USAID, Department of Justice, and a non-governmental organization. The participants offered divergent views on the underlying causes of the conflict. Both of these views of the operational environment were modeled and as whole-of-government plans were developed, they were simulated in both of these modeled environments to see the full range of possible effects.

The main body of the experiment was a two-week event in October 2007 that included ten staff with civilian agency experience and fifteen with military experience. During the first week, participants trained in using the COMPOEX tool suite. The second week began with the development of plans to achieve ten effects by the three lines of effort teams: 1) Reconstruction, 2) Governance, and 3) Security. Then the three individual plans were combined, and multiple simulations performed to determine the best use of available resources, eliminate duplicate actions, and minimize the negative impact of actions in other lines of effort. Hundreds (about 200-400) of actions were integrated into a comprehensive plan that achieved the required ten effects. A significant synergistic effect was seen as security, economic, governance and strategic communications plans were integrated and refined in an area. Experimentation demonstrated that interagency planning, utilizing advanced simulations and tools, could produce comprehensive whole-of-government plans with supporting analysis faster, in more depth, than the traditional planning tools and methods (see the Experimental Results section).

# Interagency Concept of Operation during Experiments

In our experiments, the concept of operation largely evolved as the interagency teams devised their own process. Partly due to the influence of military members, the teams largely gravitated toward a planning process reminiscent of the standard military decision making process, a formal seven-step process (US Army 1998). Typically, the core elements of this interagency process that focused on planning teamwork were (Fig. 2):

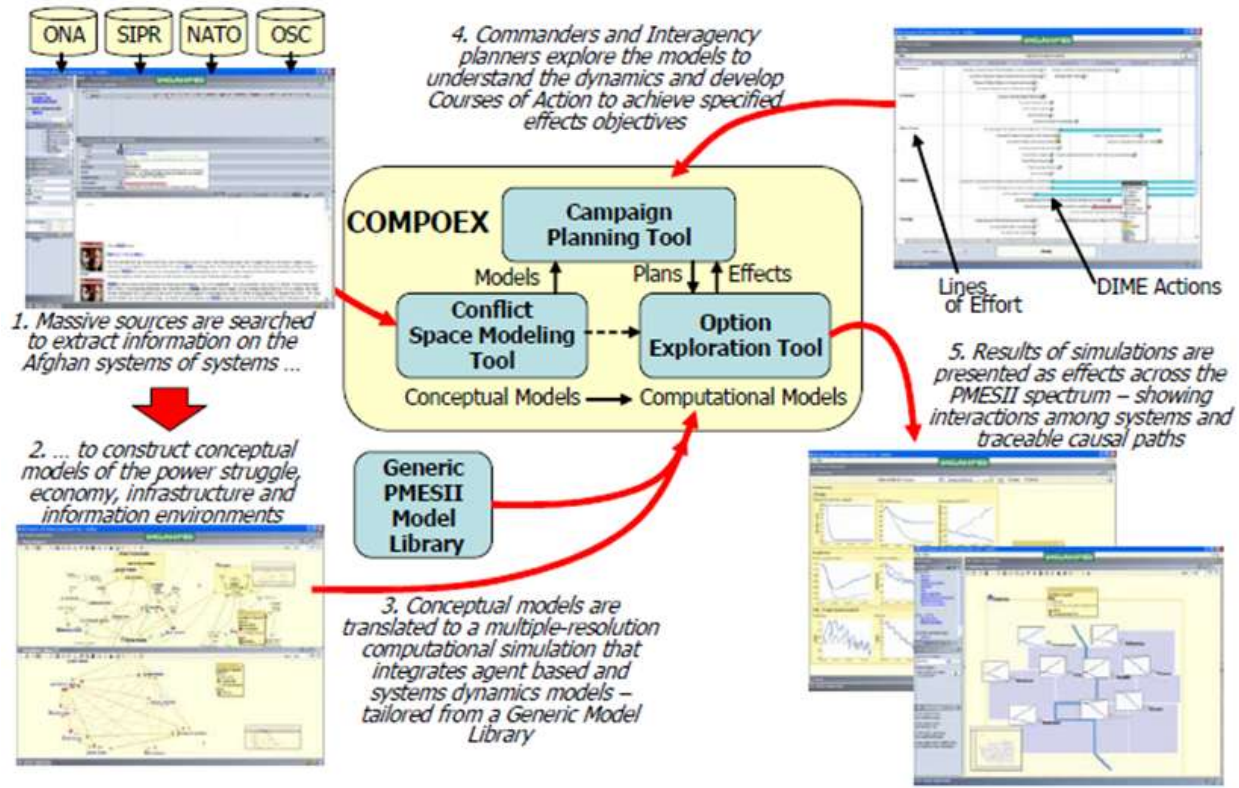

*Fig. 2. Experiments involved intertwined processes of situation analysis, planning, wargaming and assessment.*

- *Assessment: Situation Analysis* – The initial activity is the development of shared understanding of the situation (a theory of the situation or conflict) that addresses hypotheses of the underlying causes of conflict, tension, or instability. The interagency team must search databases, display information, and develop conceptual models (textual, graphical, or numerical) so the leaders can consider various concepts of the conflict. There may be alternative agency positions (competing hypotheses) on the key centers of power, leaders, the connections between the key elements, and even the underlying causes of conflict. One or more alternative conceptual models of the situation are then instantiated in a baseline computational model of the political, social, military, economic, information and infrastructure aspects.

- *Planning: Development of Candidate Course of Action (COAs)* – Planners develop an integrated operation using tools that allow individual actions, their parameters and durations to be laid out in a logical sequence with all of the dependencies between actions identified. The graphical display allows the staff to see disconnects and manipulate the timing of actions for maximum effect. Thus different components of the plan are developed separately and then brought together and refined. An integrated plan is used for execution, but alternate plan approaches and assumptions are maintained.

- *Wargaming* : *Exploration of Effects* – The plan is simulated within the virtual situation (and against alternative versions of the situation). The simulation tools are used in two ways: 1) component plan elements (e.g., the State Department governance element, a USAID humanitarian development effort, or a DoD Security element) may be simulated to understand the *individual* effects of each component on systems within the situation (e.g. governance, humanitarian aid and security effects on corruption), 2) integrated plans that combine all elements to understand the interacting effects across systems.

- *Assessment: COA Analysis and Comparison* – The effect of simulations is to help each agency understand their direct effects and interactions with other systems that produce indirect effects; this also compares effects of alternative plans. The result is plan refinement and comparison of alternative approaches.

Beyond this planning process, the concept of operations (CONOP) allows the staff to evaluate planned effects of actions against actual ongoing results on the ground. Metrics that describe each next state (Kott et al. 2007) of progress on the ground are linked to the effects and parameters expected by the models in the virtual situation. As the operation unfolds, progress is measured against what the plan expected, allowing interagency leaders and their staffs to assess progress and modify the plan, reallocate resources or modify desired effects. At the same time, they can reassess the underlying theory of the situation and assumptions to see if models need to be refined to represent reality.

## Tools and Environment of the Experiments

The human participants included a team of experienced planners from diverse agencies (primarily DoD, State, and USAID) and senior leaders (general officers, an ambassador, and National Security Council officials). The tool environment evaluated was the COMPOEX (Waltz 2009a, 2009b).

COMPOEX is a client-server system, allowing 25 planners to simultaneously assess situations, develop plans and run simulations to explore effects. The interagency teams were often organized into five planning cells, with five to six persons in each cell; the cells were organized by line of effort, for example: governance, security, economic, humanitarian, etc. For the experiments, each planning cell had five laptops and all 25 client laptops (5 cells x 5 laptops) were networked to the central COMPOEX server that allows concurrent development of component plans, integration of plan components and running the exploratory simulations. The COMPOEX toolset is comprised of several integrated elements (Fig. 3):

- Conflict Space Modeling Tools – Provides the capability to search data sources (e.g., open sources, secret internet, special holdings), capture relevant PMESII data, and construct graphical conceptual models of PMESII systems. Political-social-military network models are diagrammed as networks; economic infrastructure and information systems are diagrammed as systems flows. These conceptual representations are then translated to computational models (Fig. 3) by adapting a library of generic PMESII system model components, tailoring model parameters and structures.

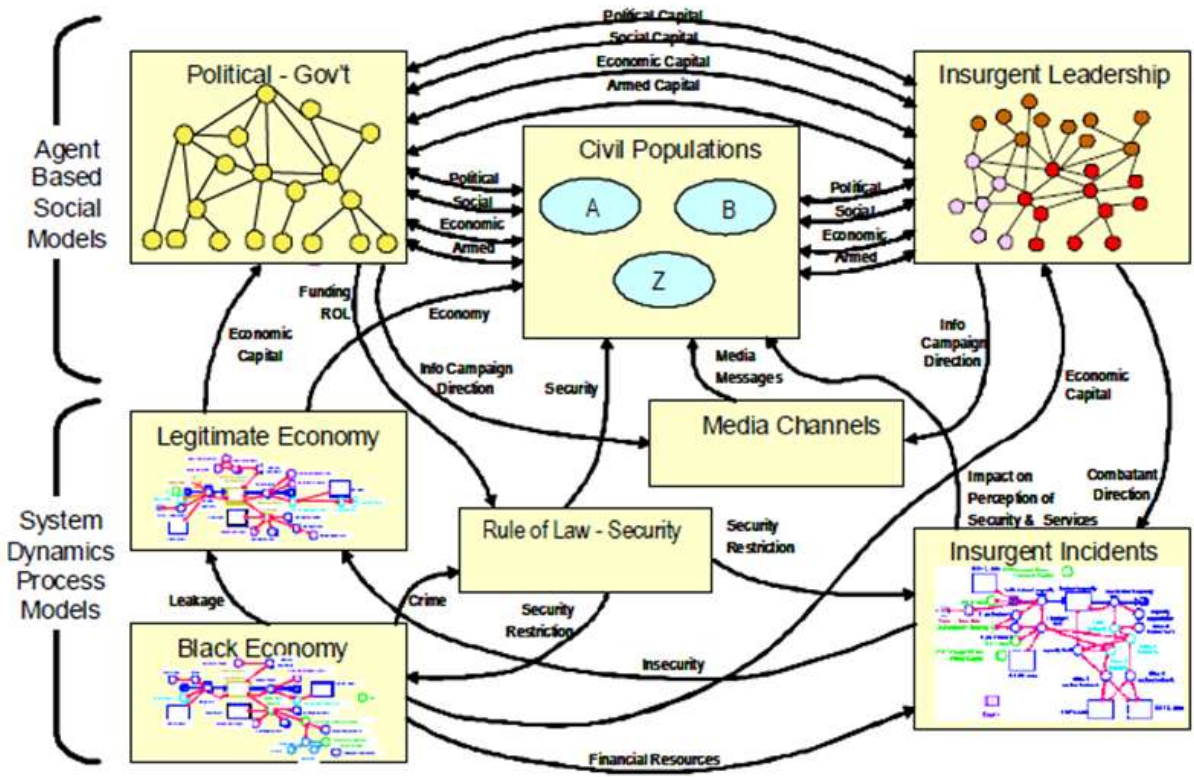

*Fig. 3. Computational models reflect the diversity of PMESII phenomena, and the range of capabilities of multiple agencies.*

- Option Exploration Tool – The collection of PMESII model components is composed into an integrated multi-resolution model (MRM) that can simulate a baseline of future behavior (e.g., stagnant growth, increasing corruption, expanded terrorist influence and unrest), and the effects from candidate US and coalition actions. The tool allows planners to explore the behavior of systems within the MRM and evaluate specific effects of optional sequences of actions.
- Campaign Planning Tool – Allows planners to schedule coordinated diplomatic, information, military and economic (DIME) actions along multiple lines of effort categories (e.g. economic, governance, strategic communications, etc.) in a synchronization matrix format. The planner enters the attributes unique to each discrete action (e.g., time of economic action start, action duration, rate of investment, source of investment, targeted economic sectors, targeted geographic region or population, etc.) and the resources required (e.g., financial resources, personnel, etc.).

# Experimental Methods, Data Collection and Reduction

In the interest of brevity, we focus here on Experiment 4. The theoretical framework of distributed cognition (Hutchins 1995) was used to analyze and assess the socio-technical system of COMPOEX for interagency planning and coordination. Distributed cognition emphasizes the distributed nature of cognitive phenomena across individuals, tools/technologies, and internal/external representations (e.g., Hansberger 2008). As with many socio-technical systems, COMPOEX (tools and human interagency team) possessed several distributed cognitive attributes including: 1) mental models, 2) workload management, and 3) coordination across agents.

Briefly, the mental model attribute refers to the representation of knowledge and its network of relationships built over time to help guide and direct behavior and decision-making. The workload management attribute focuses on the level of workload for a task or series of tasks and the factors that may affect this workload: strategies, organizational structures, and standard operating procedures. The coordination attribute addresses person-to-person and person-to-artifact interactions within the task environment.

A variety of methods were used to collect data against the distributed cognitive attributes of mental models, workload assessment, and coordination across agents (Hansberger, Schreiber, and Spain 2008). Traditional performance measures and outcomes were also collected. Complementing performance measures with the examination of distributed cognitive attributes goes beyond measuring what effect COMPOEX had on interagency planning but addresses <u>why</u> COMPOEX had the effects it did.

<u>Mental Models.</u> Mental models have a long history in Psychology and Cognitive Science (e.g., Johnson-Laird 1983) as the cognitive representation of accumulated knowledge and experience. One established way to measure mental models is through the measurement of structural knowledge, which is the pattern of relationships between concepts in declarative memory. These concepts have varying degrees of interrelatedness with each other where some are more closely related to the targeted concept than others. In order to assess these relationships and the varying strengths of them, individuals can rate the similarity between concepts (Jonassen et al. 1993). The representation of the knowledge structures elicited by the above similarity ratings can be accomplished through a network approach using Pathfinder software and Pathfinder networks (Schvaneveldt 1990). Pathfinder uses the pair wise proximity estimates for a set of concepts and generates a network structure where the concepts are nodes and the relations between concepts are links in the network structure (Fig. 4).

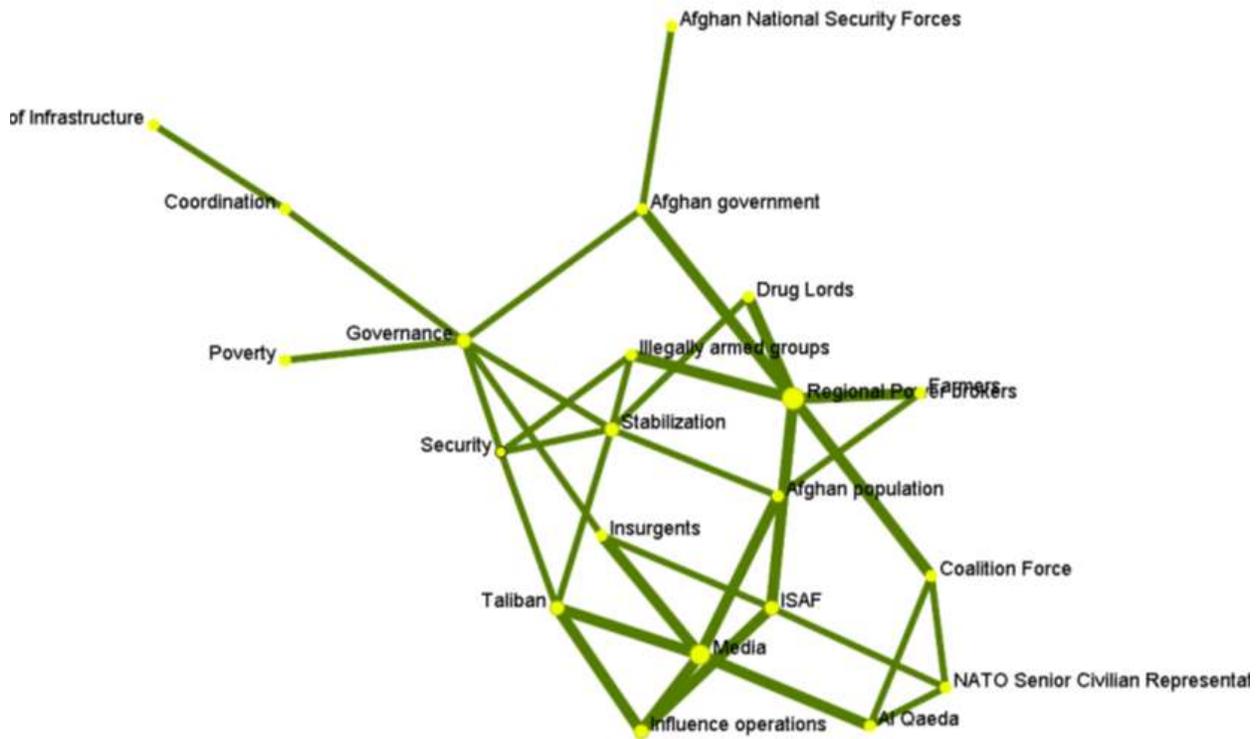

*Figure 4. Expert knowledge structure created using the Pathfinder method.*

Workload Assessment. The NASA Task Load Index (TLX) (Hart and Staveland 1988) is a subjective workload assessment measure that allows users to perform subjective workload assessments on operator(s) working with various human-machine systems. TLX is a multi-dimensional rating procedure that derives an overall workload score based on a weighted average of ratings on six subscales. It can be completed in a short amount of time through a simple computer program.

Coordination Across Agents. Social network analysis (SNA) uses graphs as a representation of symmetric or asymmetric relations between discrete objects (Scott 2000). Placed within a social context of humans and their interactions, a social network is a set of individuals (i.e., nodes) connected through social interactions like face-to-face or email communication (i.e., links). Person-to-person coordination was collected using an observational data collection tool developed by the Army Research Laboratory (Hansberger, Schreiber, and Spain 2008) called SNA Observer. This tool allowed the observers to document all team interactions regarding who talked with whom and the duration of that interaction. This data was then available for analysis using social network analyses described below. Person-to-system or tool interactions were also collected using the SNA Observer as well as internal computer log data.

# Discussion of the Experimental Results

The experimental results for experiments 1-3 will be summarized briefly due to space limitations and experiment 4 will be discussed in further detail.

Experiment 1: Effects Identification. We found that interagency teams equipped with the COMPOEX tools were able to identify a significantly greater number of important PMESII effects as compared to a control team that operated in a conventional, manual fashion. For example, depending on the scenario, the control team was able to identify 10 to 44 potential unfavorable effects of an operation plan; while teams with tools identified up to 313 significant effects across the variety of PMESII systems.

Experiment 2: Domain Visualization. This experiment focused on the effect the COMPOEX visualizations had on interagency planners' mental models and understanding of the relationships. Measuring mental models of participants, we found that the visualizations were effective enough to allow the less experienced planners to answer questions at the same level as more experienced planners.

Experiment 3: Operation Planning. Among the distributed cognitive attributes, participant mental models, workload management, and coordination across agents were measured. The results in the areas of planning, setting objectives and alternatives, and mental models showed that the Campaign Planning Tool was as effective, if not more effective, in facilitating the planning process compared to a team using more traditional tools and methods. The results from the area of data manipulation showed that there was no increase in workload across any of the six dimensions measured.

Experiment 4: Parallel Planning. The fourth experiment investigated the use of the models and the ability to explore options within the plan. Several planning performance measures were examined along with the distributed cognitive attributes of mental models, their change over time, understanding and coordination across agents using SNA, and measures of trust in the COMPOEX tools and simulations.

<u>Mental Models.</u> The participants' overall understanding, when their Pathfinder knowledge structures were compared with a subject matter expert (SME), showed a significant positive trend over time ($R^2 = .89$, $p = .05$) (Fig. 5). This increase over time suggests that the COMPOEX tools had a positive influence on improving understanding among the users. The initial knowledge state of the COMPOEX users and a comparative planning cell using current procedures and technology measured about the same when compared to the SME Pathfinder knowledge structure. However, as the interagency planners interacted with COMPOEX, they significantly increased their knowledge over time (Fig. 5).

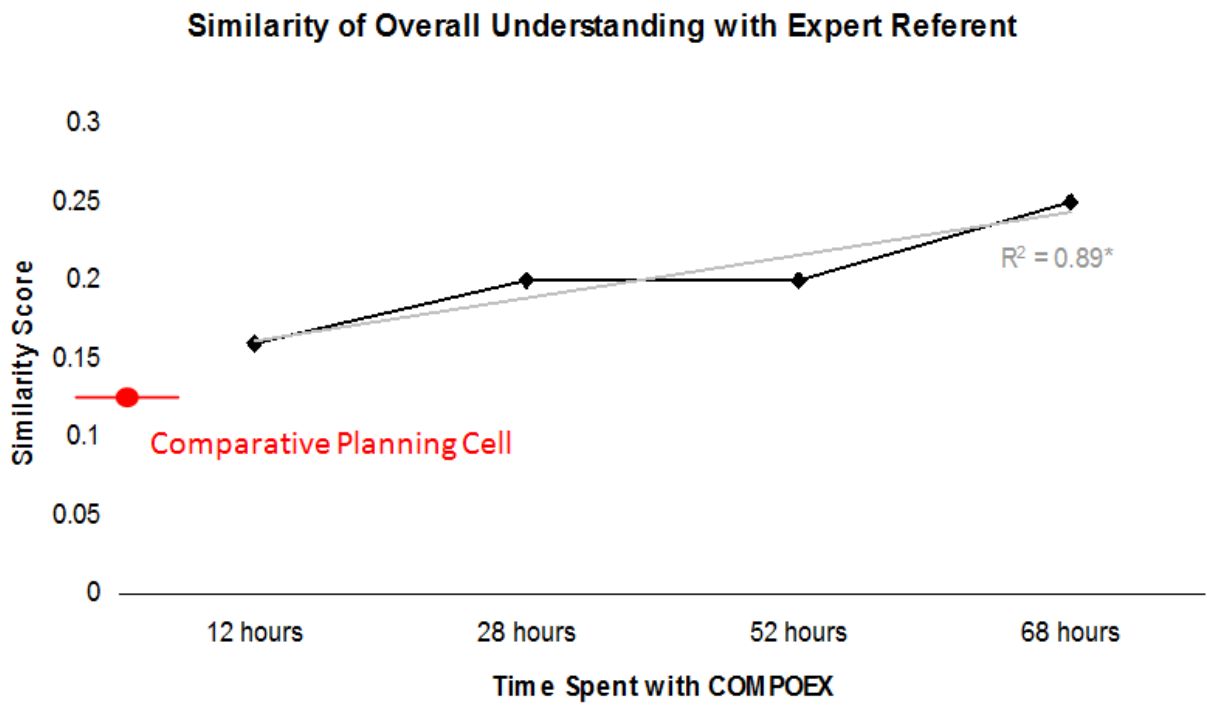

*Fig. 5. Understanding over time across all participants—a significant positive trend.*

Coordination Across Agents.

Coordination within the interagency team showed that tools did not restrict collaboration across lines of effort, between leadership and planners, nor between military and civilian/government experienced participants (Fig. 6). Patterns of coordination also showed that reliance on tool-support personnel declined over time suggesting that participants were able to learn and increase their proficiency with the COMPOEX tools.

A key element for coordination within any socio-technical system is trust among the socio-technical agents (human and computer/automation). Participants completed a 13-item human-machine trust measure (a modified version of Jian, Bisantz, and Drury's (2000) scale) to report the level of trust they maintained in the MRMs during the exercise. Trust ratings clustered around the mid-point of the scale, thus indicating that participants did not over-trust; nor distrust the MRMs.

This experiment also examined whether participants would use the MRMs for explanatory or predictive purposes. First, perceptions of predictive value were compared to perceptions of explorative value. As shown in Fig. 7, participants believed that the MRMs had more explorative than predictive value ($t = -11.9$, $p < .01$). It is worth noting that even though the explanatory score is much higher than the prediction score, the prediction score is right at the neutral score of "4" and is not significantly lower. This suggests that participants used the MRMs for prediction purposes but not to the degree they used them for explanation. In addition, military and civilian planners had the same level of trust in the tools—an important issue for interagency teams.

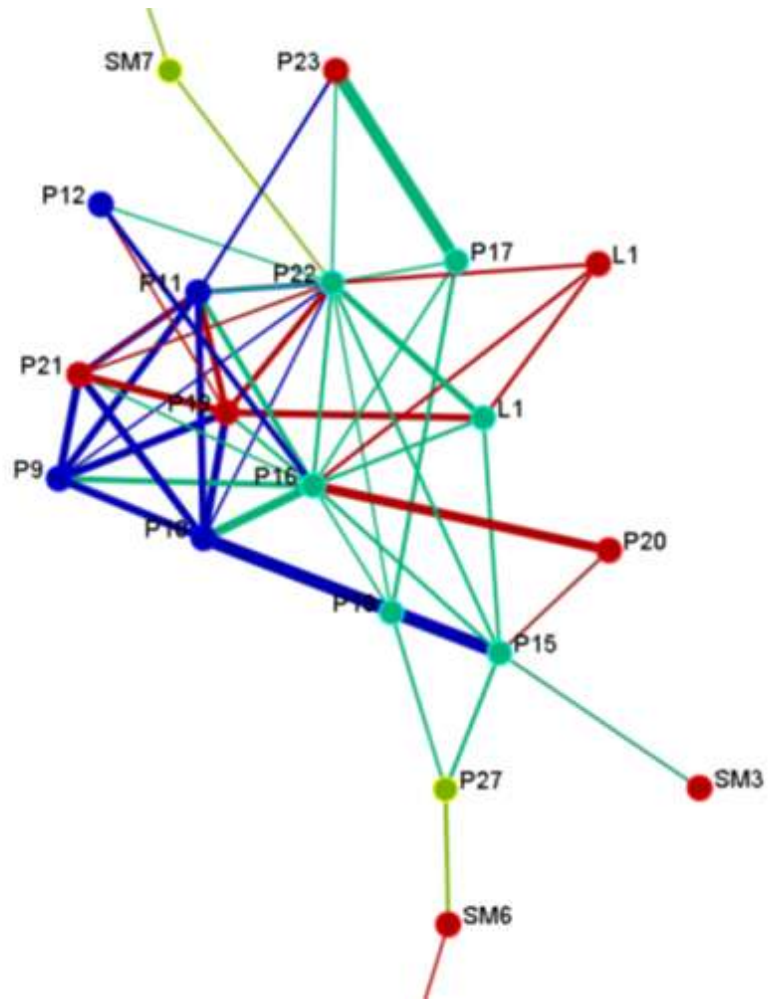

*Fig. 6. Patterns of communication for interagency planners. Governance members are teal, reconstruction is red, security is blue, and yellow-green is the strategic cell.*

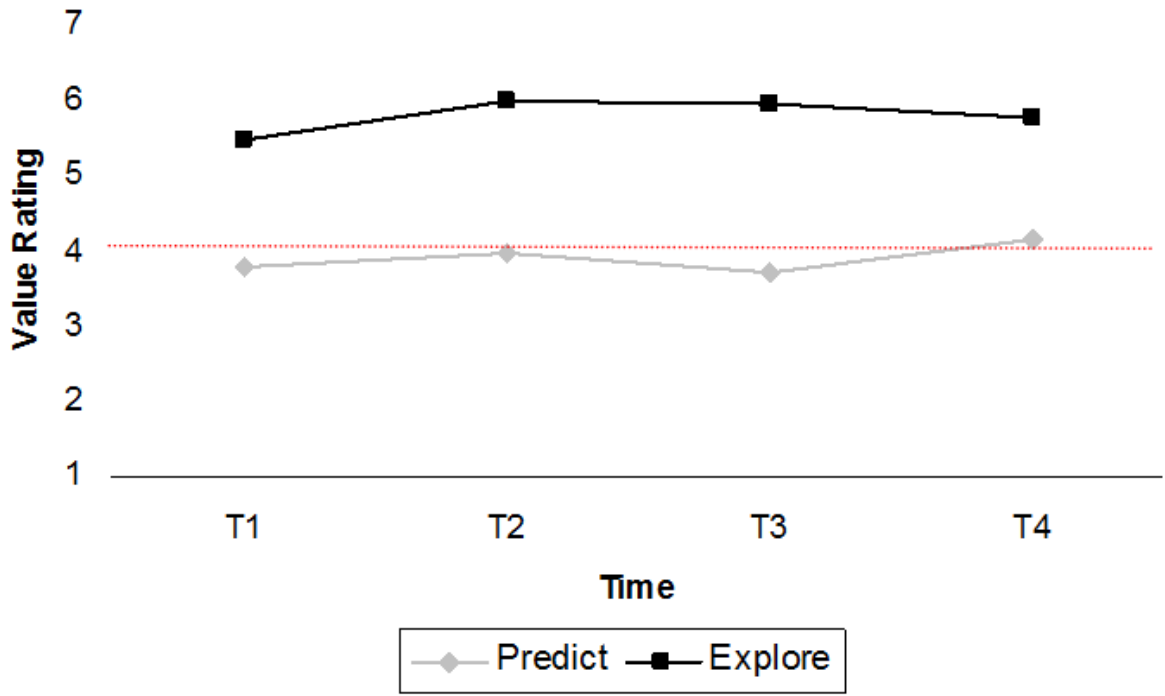

*Fig. 7. Comparison of explorative value vs. predictive value.*

Performance results. The output and products that COMPOEX facilitated were evaluated in terms of the time and effort to develop the MRMs and the planning outputs produced. The time and resources required compared favorably with the methods largely used today. Hundreds of actions (over 400) were examined in less than ten days of planning and 12 effects were integrated into a single plan. A comparable planning effort and staff unassisted by the COMPOEX system examined 150 actions and developed five effects over more than two months. Participants also perceived several benefits in the use of COMPOEX over existing methods and tools; e.g., the ability to explore individual, multiple, and combined actions in the context of a complex interagency planning effort.

## Lessons Learned and Recommendations

The interagency nature of such experimentation introduces special practical and methodological challenges, and requires careful attention.

To begin with, organizers of an interagency experiment must recognize that some agencies are significantly less well equipped to participate in interagency experiments or any experiments at all. Generally less experienced in experimentation, they often have difficulties appreciating the importance of experiments and releasing personnel from on-going real world operations.

Pre-experiment training is a strict necessity. An interagency team has to dedicate several days and even weeks to learn how to operate jointly as a team, to formulate and rehearse a concept of operation (often entirely unfamiliar and uncomfortable to some members), and to adjust to vocabulary, concepts and conventions of members from other agencies.

When performing the series of related experiments, a campaign of experiments that takes several years to complete, as in our case, the experiment organizers face changes in priorities and requirements of not one but multiple agencies. To maintain the buy-in and participation of multiple agencies-stakeholders, the experimenters invest significantly greater time in coordination and relation maintenance. Further, the design of experiments has to evolve and adapt to meet the changing (and potentially conflicting) interests of multiple agencies.

Such additional complications – procurement of interagency personnel, extra logistics, coordination, training, etc. – make interagency experiments more expensive. The experiment organizers must plan for the inevitable additional expenses.

The experiment design must employ knowledge and sensitivity to individual inter-agency cultures and climates. Similar to Hofstede's (2001) analysis across nationalities, the critical similarities and differences can be investigated through a variety of means ranging from an extensive experimental approach to interviews with members of that agency/organization.

The design of experiment must take into account the broad range of multiple agencies' interests. To over-simplify, one agency may be concerned with loss exchange ratio while another with per-capita food production. Such diverse interests must be taken into account while devising control conditions, independent and dependent variables, data collection and processing techniques, and overall metrics.

In particular, the composition of an interagency team and interagency experiences of individual members--is a highly influential independent variable that is difficult to control. For example, because military members usually have significant experience and specific training in command and control processes, they may heavily influence civilian members of the team.

It is important to understand the patterns of interactions and communications across inter-agency members and organizations. The dynamics and development of these interactions over time can play as critical a role in inter-agency performance and success as any other variable (Cross 2004). There is a variety of collection and analysis means ranging from observational data collection to varying levels of automated means.

Finally, of particular importance in interagency experimentation are the dynamics and effects of distributed cognition that occurs in an interagency team. Command and control processes, such as planning, often have the emergence of a common perception and vision as most important outcomes of the process. Thus, interagency experiments should pay particular attention to attributes that reflect dynamics of mental models and coordination within the heterogeneous interagency team.